\documentclass{appolb}
\usepackage{graphicx}
\usepackage{amssymb}
\usepackage{hyperref}
\usepackage{color}
\usepackage{xurl}
\usepackage{xspace}
\newcommand{\pt}{\mbox{$p_T$}\xspace}

\newcommand{\meanncoll}{\mbox{$\langle N_{\mathrm coll} \rangle$}\xspace}

\newcommand{\pp}{\mbox{$p$$+$$p$}\xspace}

\newcommand{\dau}{\mbox{$d$$+$Au}\xspace}

\newcommand{\pau}{\mbox{$p$$+$Au}\xspace}
\newcommand{\pal}{\mbox{$p$$+$Al}\xspace}

\newcommand{\jpsi}{\mbox{$J/\psi$}\xspace}
\newcommand{\psip}{\mbox{$\psi$(2S)}\xspace}

\definecolor{cyan}{rgb}{0.0, 1.0, 1.0}
\definecolor{airforceblue}{rgb}{0.36, 0.54, 0.66}

\hypersetup{
colorlinks=true,
    linkcolor={airforceblue},
    filecolor={airforceblue},      
    urlcolor={airforceblue},
    citecolor = {airforceblue},
    pdftitle={airforceblue},
        }

\begin{document}
\title{J/$\psi$ and $\psi$(2S) Production in Small Systems with PHENIX%
\thanks{Presented at the 29$^{th}$ International Conference on Ultrarelativistic Nucleus-Nucleus Collisions (Quark Matter 2022)}%
}
\author{Krista Smith, Los Alamos National Laboratory\\
(for the PHENIX Collaboration, \\
\href{https://urldefense.com/v3/__https://doi.org/10.5281/zenodo.7430208__;!!Epnw_ITfSMW4!qgbQcPCW04Dksv85IroFOUtyJjpToohQBsccKLOfRaG4b-yXl6hbrqw5W_z1RiJRCWVzpyZdvnnIh2YbKQ$}{\tiny{\url{https://zenodo.org/record/7430208}}}) }

\address{}
%\\[3mm]
{ % of different affiliation
\address{}
}
%\\[3mm]
%the Name(s) of other Author(s)
\address{}

\maketitle
\begin{abstract}
The suppression of the $\psi$(2S) nuclear modification factor has been seen as a trademark signature of final state effects in large collision systems for decades. In small systems, deviations of the nuclear modification from unity had been attributed to cold nuclear matter effects until the observation of strong differential suppression of the $\psi$(2S) state in $p/d+$A collisions, which suggests the presence of final state effects. In this paper, we present results of J/$\psi$ and $\psi$(2S) measurements in the dimuon decay channel for $p+p$, $p+$Al, and $p+$Au collision systems at $\sqrt{s_{NN}}$ = 200 GeV. Key results include the nuclear modification factors $R_{pA}$ as a function of centrality and rapidity. The measurements are compared with shadowing and transport model predictions, as well as to complementary measurements at LHC energies.
\end{abstract}
  %%==============================
  % 6 page limit
  %===============================
\section{Introduction}
We present the first results of PHENIX \psip~nuclear modification measurements at forward and backward rapidity.  Here we discuss a selected number of results from a recent PHENIX publication~\cite{PHENIX:2022nrm} that looked for signs of final-state effects on charmonium production in small system collisions at RHIC energies. 

\section{Experimental Overview \& Data Analysis} % DONE
The PHENIX Muon Arms are located parallel to the beam pipe, and measure muons and unidentified charged hadrons at forward (1.2 $< \eta < 2.4$) and backward ($-2.2 < \eta < -1.2$) rapidity.  The Muon Arms compromise four main components:  the FVTX Detector, the Muon Tracker, the Muon Identifier, and a series of steel hadron absorbers located throughout each Muon Arm.   
The measurements were attainable using the Forward Vertex Silicon Detector (FVTX), installed in the PHENIX 2012 upgrade, which covers the pseudorapidity range $1.2 < |\eta| < 2.2$. The FVTX detector, a precision silicon tracking detector, provides additional space points closest to the interaction region, supplying the mass resolution necessary to extract the \psip~signal.  

%======================================
%  Figure 1 Venn ( RHIC thesis award tpsalk )
\begin{figure}[t!]
\centerline{%
\includegraphics[width=7.5cm]{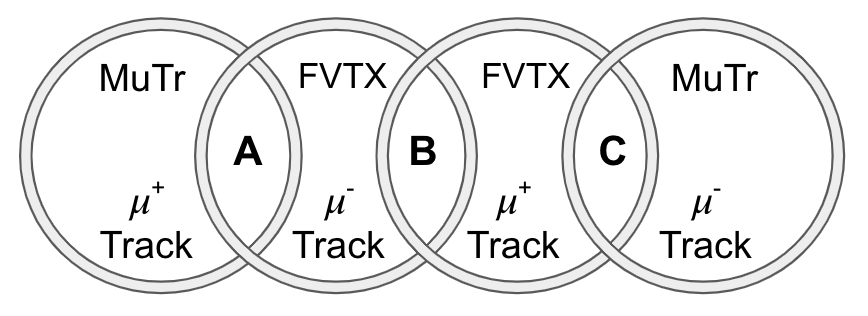}}
\caption{Venn diagram illustrating the different combinations of reconstructed dimuon pairs used to increase the statistics available for the \psip~measurements while still utilizing the mass resolution of the FVTX detector.}
\label{Fig:Fig1}
\end{figure}
%======================================

%\section{Data Analysis}  % DONE

The \psip~measurements are presented in three small system data sets at the center of mass energy $\sqrt{s_{_{NN}}}$ = 200~GeV: \pp, \pal, and \pau~collisions, all recorded during the 2015 run year. Measurements taken at forward rapidity denote the $p$-going direction of the collision, while backward rapidity denotes the Al/Au-going direction. 
In this analysis, the FVTX detector was used in conjunction with the Muon Tracker and the Muon Identifier to reconstruct the dimuon invariant mass. Previous PHENIX \psip~measurements~\cite{PHENIX:2016vmz} required both tracks in the reconstructed dimuon pairs to be within the FVTX acceptance region. This requirement significantly improves the reconstructed dimuon mass resolution, but at the expense of statistics due to the limited geometric acceptance of the FVTX detector.

A new analysis method was developed to extract the \psip~nuclear modification measurement in the PHENIX Muon Arms. This method combines muon tracks inside and outside the FVTX acceptance region, increasing the available statistics (see Figure~\ref{Fig:Fig1}). Dimuon pairs are reconstructed in two different ways:  both muon tracks are within the FVTX acceptance (denoted by Intersection B of Figure~\ref{Fig:Fig1}), or just one muon track is inside the FVTX acceptance (denoted by Intersections A and C).

\section{Results}

%======================================
%  Figure 3 ( ppg245 resub PRC )
\begin{figure}[t!]
\centerline{%
\includegraphics[width=8cm]{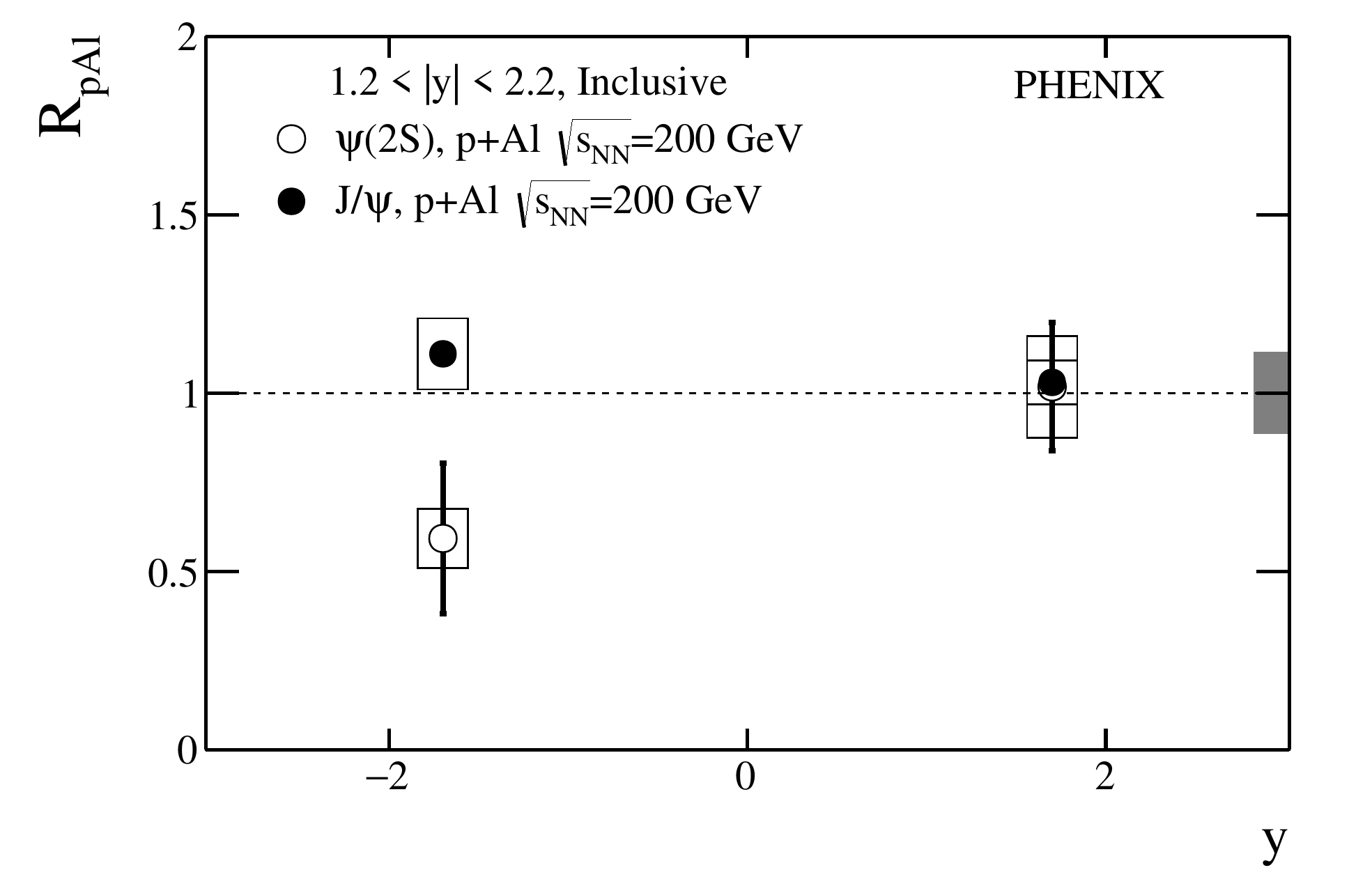}}
\caption{\label{fig:RpAl_rap} The \pt~and centrality-integrated nuclear modification factors as a function of rapidity for \psip~[open circles] and \jpsi~[solid circles] in \pal~collisions.}
\end{figure}
%======================================
%===================================
%  Figure 4 ( ppg245 resub PRC )
\begin{figure}[t!]
\centerline{%
\includegraphics[width=8cm]{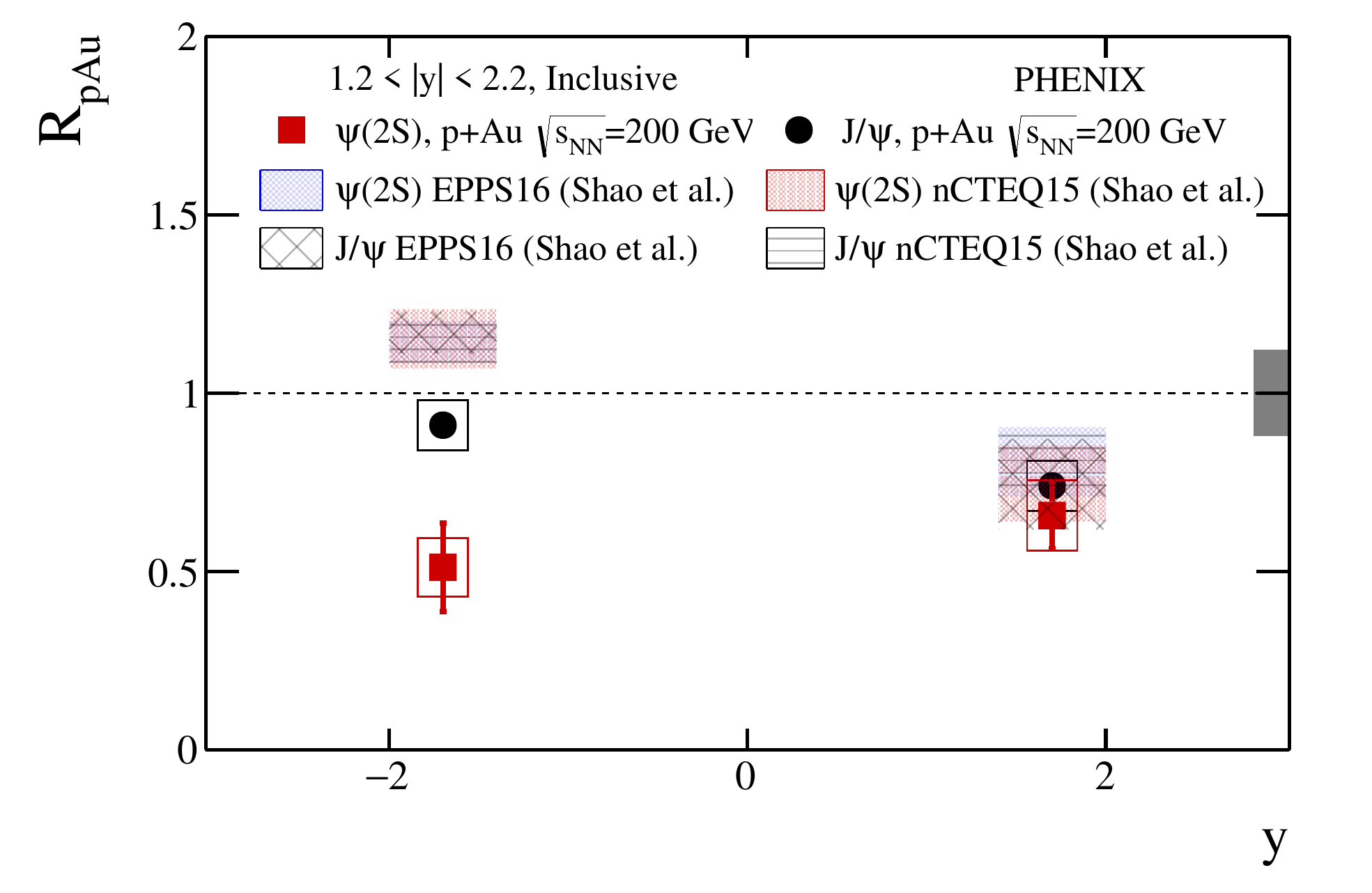}}
\caption{
\label{Fig:FigRpAu} The \pt~and centrality-integrated nuclear modification factors as a function of rapidity for \psip~[red squares] and \jpsi~[black circles] in \pau~collisions.  The data is compared to EPPS16 and nCTEQ15 shadowing predictions.}
\end{figure}
%======================================

% Results 1
%=======================================
Shown in Figure~\ref{fig:RpAl_rap} is the \psip~(open black data points) and \jpsi~(solid black) nuclear modification factor in \pal~collisions.  The measurements correspond to \pt $>$ 0 GeV/c, and are taken over the full rapidity range of $1.2 < |y| < 2.2$.  At forward rapidity, the nuclear modification measurements are consistent between the two states.  There is little to no nuclear modification, as would be expected from a lighter Aluminum target.  As observed in a similar PHENIX analysis \cite{PHENIX:2019brm}, \jpsi~nuclear modification measurements did not show any effects of gluon shadowing at forward rapidity.  

At backward rapidity, the \jpsi~nuclear modification measurement is consistent with unity.  This implies that gluon anti-shadowing and nuclear absorption effects, if any, are small.  Therefore the suppression seen in the \psip~nuclear modification factor at backward rapidity cannot be attributed to nuclear absorption.  This suppression is possibly due to final state effects, but the error bars are large and do not permit a clear conclusion.

% Results 2
%=======================================
A complementary measurement in the \pau~collision system is shown in Figure~\ref{Fig:FigRpAu}.  The \psip~nuclear modification measurements (red squares) are compared with the \jpsi~nuclear modification measurements (black circles).  Again, the modification between the two states is consistent at forward rapidity.  The data is well described by EPPS16~\cite{Eskola:2016oht} and nCTEQ15~\cite{Kovarik:2015cma} re-weighted gluon shadowing predictions~\cite{Kusina:2017gkz} as the source of suppression at forward rapidity.  

At backward rapidity, the \jpsi~suppression can be understood as a trade-off between the competing effects of anti-shadowing and nuclear absorption.  Cold nuclear matter effects for the \jpsi~and \psip~states are expected to be similar for measurements recorded with the same rapidity, collision system and collision energy.  Therefore nuclear absorption cannot explain the suppression of the \psip~nuclear modification with respect to the \jpsi~suppression.  Furthermore, gluon shadowing alone cannot describe the data.

% Results 3
%======================================
%  Figure 8 ( ppg245 resub PRC )
\begin{figure}[t!]
\centerline{%
\includegraphics[width=15cm]{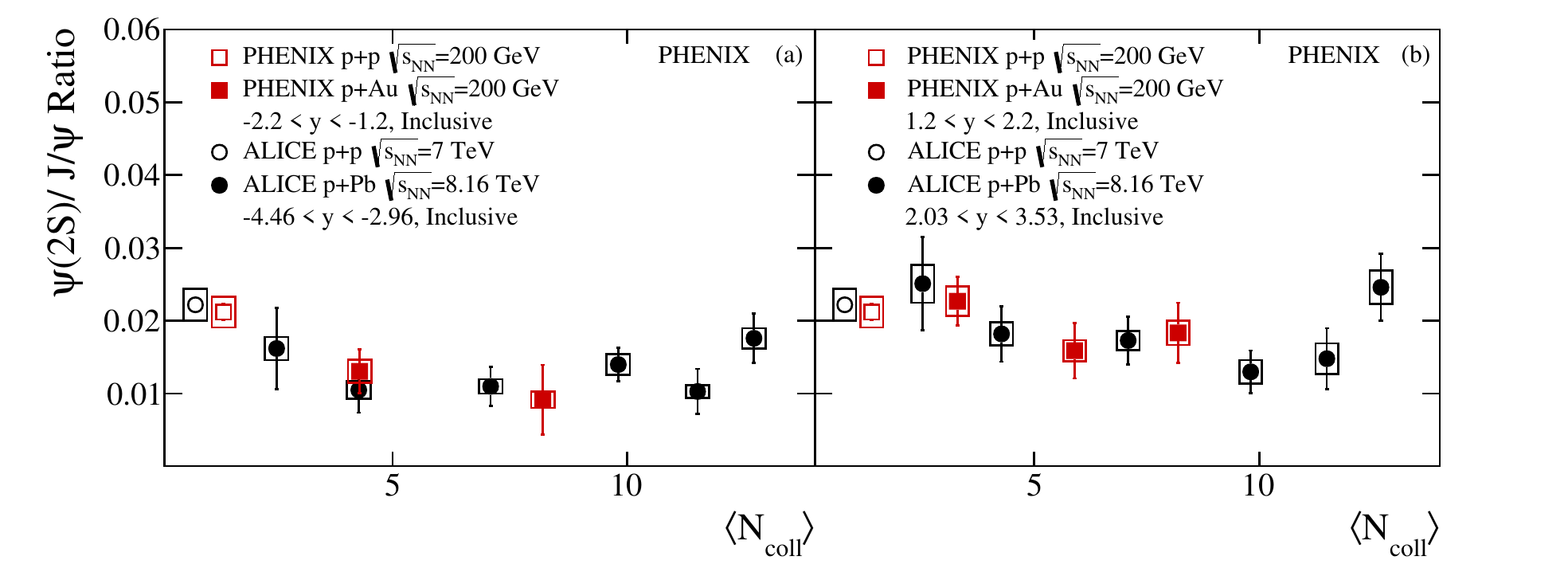}}
\caption{
\label{Fig:Fig_alice}  The \psip~to \jpsi~ratio is shown as a function of \meanncoll~at backward (a) and forward (b) rapidity.  PHENIX data in \pau~collisions compared to ALICE data in $p$+Pb collisions.  The corresponding ratios in \pp~collisions for each experiment are shown for \meanncoll~= 1.}
\end{figure}
%======================================

Figure~\ref{Fig:Fig_alice} shows the \psip~to \jpsi~ratio as a function of \meanncoll.  The PHENIX data in \pau~collisions (solid red squares) is directly compared to ALICE data~\cite{Acharya:2020rvc} in $p$+Pb collisions (solid black circles).  The corresponding \psip~to \jpsi~ratio by PHENIX (open red square) and ALICE (open black circle) is shown for \pp~collisions at \meanncoll~= 1.  Previous publications \cite{PHENIX:2016vmz} have shown the \psip~to \jpsi~ratio in \pp~collisions is independent of collision energy.  It is expected that by taking the ratio of the two charmonium states, cold nuclear matter effects largely cancel, leaving final state effects as the dominant contribution.  Despite the different Bjorken-$x$ and $Q^2$ regions probed by the two experiments, the ratio as a function of centrality is quite similar, particularly at backward rapidity.  It is worth noting that the Lorentz factor is larger at LHC energies, which might allow charmonium to escape the interaction region more quickly than at RHIC energies. 

\begin{figure}[t!]
\centerline{%
\includegraphics[width=8cm]{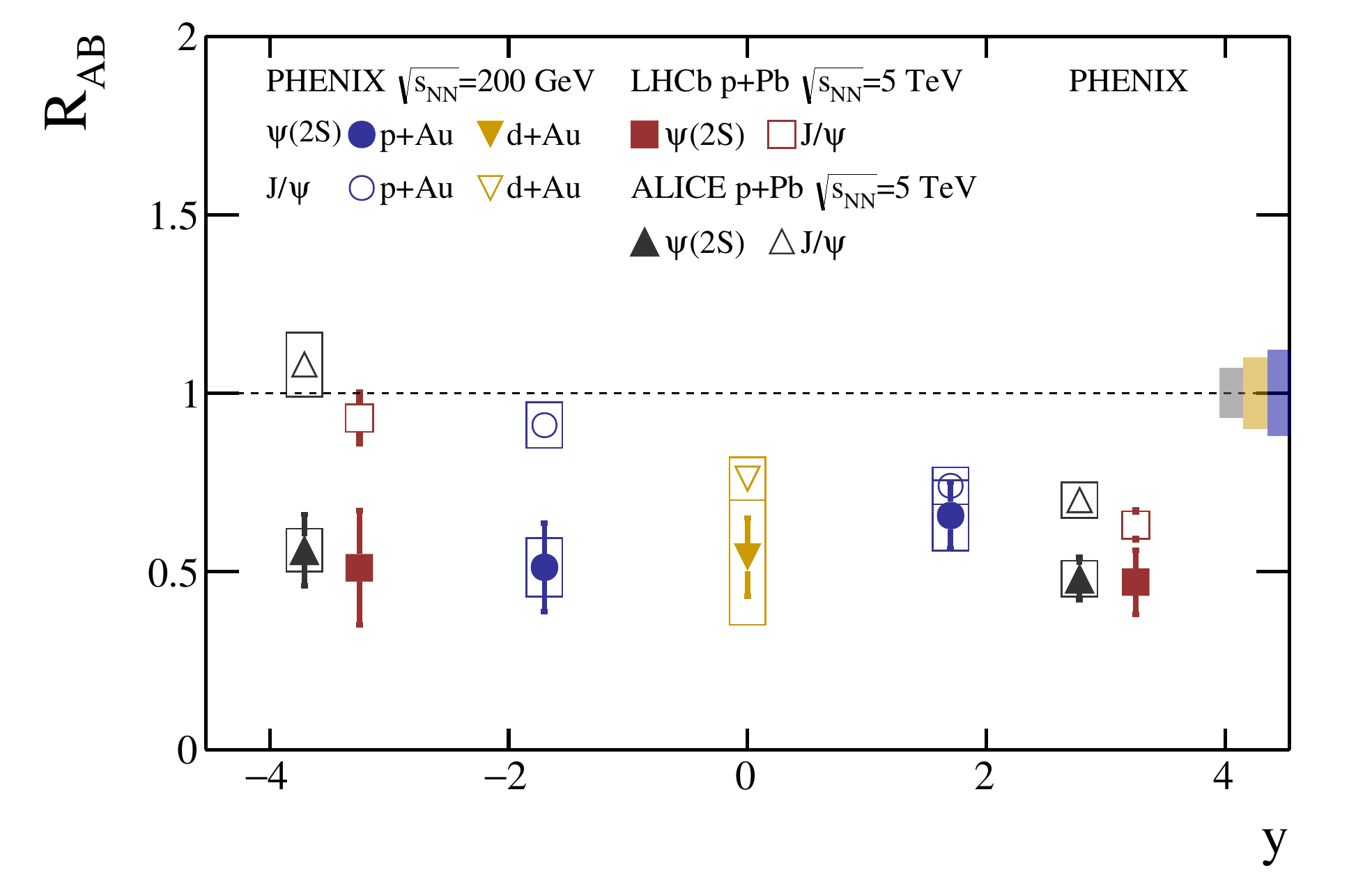}}
\caption{ \label{Fig:Fig9} Nuclear modification measurements for \psip~(solid data points) and \jpsi~(open data points) are shown for the PHENIX, LHCb, and ALICE experiments as a function of rapidity.  Refer to the text for more details. }
\end{figure}

% Results 4
%=======================================
Lastly, a more comprehensive comparison of charmonium nuclear modification measurements are shown at both RHIC and LHC energies in Figure~\ref{Fig:Fig9}.  The \jpsi~(\psip) nuclear modification is denoted with the open (solid) data points.  The measurements are taken over a variety of small collision systems, including RHIC \pau~and \dau~collisions at $\sqrt{s_{NN}}$ = 200 GeV, and LHC $p$+Pb collisions at $\sqrt{s_{NN}}$ = 5 TeV.  The current PHENIX \pau~results are shown in blue, previous PHENIX \dau~measurements in gold~\cite{PHENIX:2013pmn}, and LHCb~\cite{LHCb:2016vqr, LHCb:2013gmv} and ALICE~\cite{ALICE:2013snh,ALICE:2014cgk} in red and gray, respectively.  At forward rapidity, similar modification is seen between \jpsi~and \psip, indicating cold nuclear matter effects are dominant.  At backward rapidity, stronger \psip~suppression is seen across all three experiments, strongly suggesting final state effects are present in small collision systems.

%======================================
%  Figure 9 ( ppg245 resub PRC )
%======================================

\section{Conclusion}
We have presented \pt-integrated charmonium nuclear modification measurements and the \psip~to \jpsi~ratio as a function of centrality at forward and backward rapidity, and compared PHENIX measurements with those taken at LHCb and ALICE.  The data suggests that gluon shadowing is the dominant contribution to modification at forward rapidity for both the \jpsi~and the \psip~states.  At backward rapidity, we find stronger suppression of the \psip~with respect to the \jpsi, a signature historically attributed to color screening.  Additionally, the \psip~to \jpsi~ratio is surprisingly similar between PHENIX and ALICE data, particularly at backward rapidity, suggesting final-state effects are similar at RHIC and LHC energies.  

%\clearpage

%======================================
\bibliographystyle{elsarticle-num}
\bibliography{ppg245x1.bib}
%======================================

\end{document}